\documentstyle[11pt,newpasp,twoside,epsf]{article} 
\def\edcomment#1{\iffalse\marginpar{\raggedright\sl#1\/}\else\relax\fi}
\reversemarginpar

\begin{document}
\title{Hard GRB spectra: thermal vs non--thermal emission} 
\author{Ghirlanda G., Celotti A.}
\affil{SISSA/ISAS, via Beirut 2-4, Trieste, Italy} 
\author{Ghisellini G.}  
\affil{Osservatorio
 Astronomico di Brera, via Bianchi 46, Merate, Italy.}

\begin{abstract}
 We consider the evidence for very hard low energy spectra during the
  prompt phase of Gamma Ray Bursts (GRB).  In particular we examine
  the spectral evolution of GRB 980306 together with the detailed
  analysis of some other bursts already presented in the literature
  (GRB 911118, GRB 910807, GRB 910927 and GRB 970111), and check for
  the significance of their hardness by applying different tests.  The
  hard spectra of these bursts and their evolution constrain several
  non--thermal emission models, which  appear inadequate to account
  for these cases. The extremely hard spectra at the beginning of
  their prompt emission are also compared with a black body spectral
  model: the resulting fits are remarkably good. These findings on the
  possible thermal character of the evolving spectrum and their 
  implications on the GRB physical scenario can be considered in the
  frameworks of photospheric models for a fireball which is becoming
  optically thin, and of the Compton drag model.  Both models appear
  to be qualitatively and quantitatively consistent with the found
  spectral characteristics.
\end{abstract}

\section{Introduction}
Since their discovery, the time resolved spectral analysis of GRBs has
been a testing ground--floor for the models proposed for the
$\gamma$--ray prompt emission (Ford et al. 1995; Preece et al. 2000).
Nonetheless, the mechanism(s) responsible for their emission is still
an open problem (Ghisellini 2003). Different authors (Crider et
al. 1997, 1999; Preece et al.  1998, 2002; Ghirlanda et al. 2002) have
shown that the low energy photon spectrum (typically represented by a
powerlaw $N(E)\propto E^{\alpha}$) can be harder than the limit of the
optically thin synchrotron model ($N(E)\propto E^{-2/3}$, Katz 1994;
Tavani 1996) which is the most popular mechanism proposed for
interpreting the burst emission.  Modifications of the simplest
optically thin synchrotron model (Lloyd \& Petrosian 2000, 2002;
Medvedev 2001) or variants based on Comptonization (Liang et al. 1997;
Ghisellini \& Celotti 1999; Lazzati et al. 2000) have been proposed to
solve the inconsistency between theory and observations. All these
models predict different limits for the low energy spectral hardness
and can be directly tested by the comparison with the hardest bursts
observed by BATSE.

We present the results of the time resolved spectral analysis of
5 bursts with a low energy spectral component harder than a flat
photon spectrum, i.e. $\alpha\ge 0$, for most of their duration. These
results are shown to constrain the above emission models (Ghirlanda,
Celotti \& Ghisellini 2003). 

Another interesting aspect of GRB spectra which we consider is their
possible thermal character, which is predicted by the fireball model
(Goodman 1986; Paczynski 1986). We find that the spectrum in the
initial phase of these extremely hard bursts is successfully fitted
with a thermal black body emission. The interpretations and
implications of these results are discussed (Ghirlanda et al 2003) in
the framework of a photospheric (e.g. M\'esz\'aros \& Rees 2000,
Daigne \& Mockovitch 2002) and of the Compton drag model (Lazzati et
al. 2000; Ghisellini et al. 2000).
\begin{figure}[hbt]
\plottwo{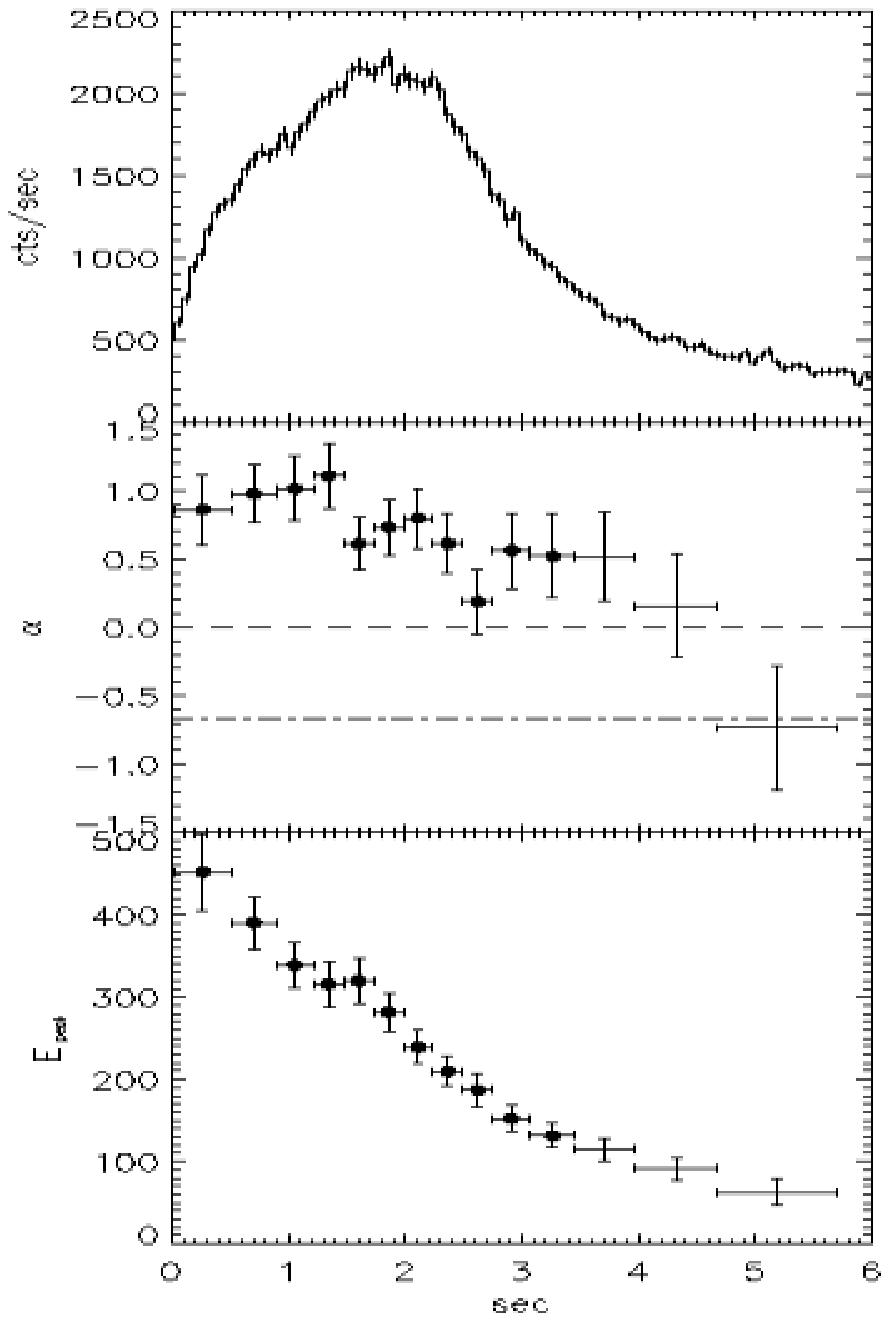}{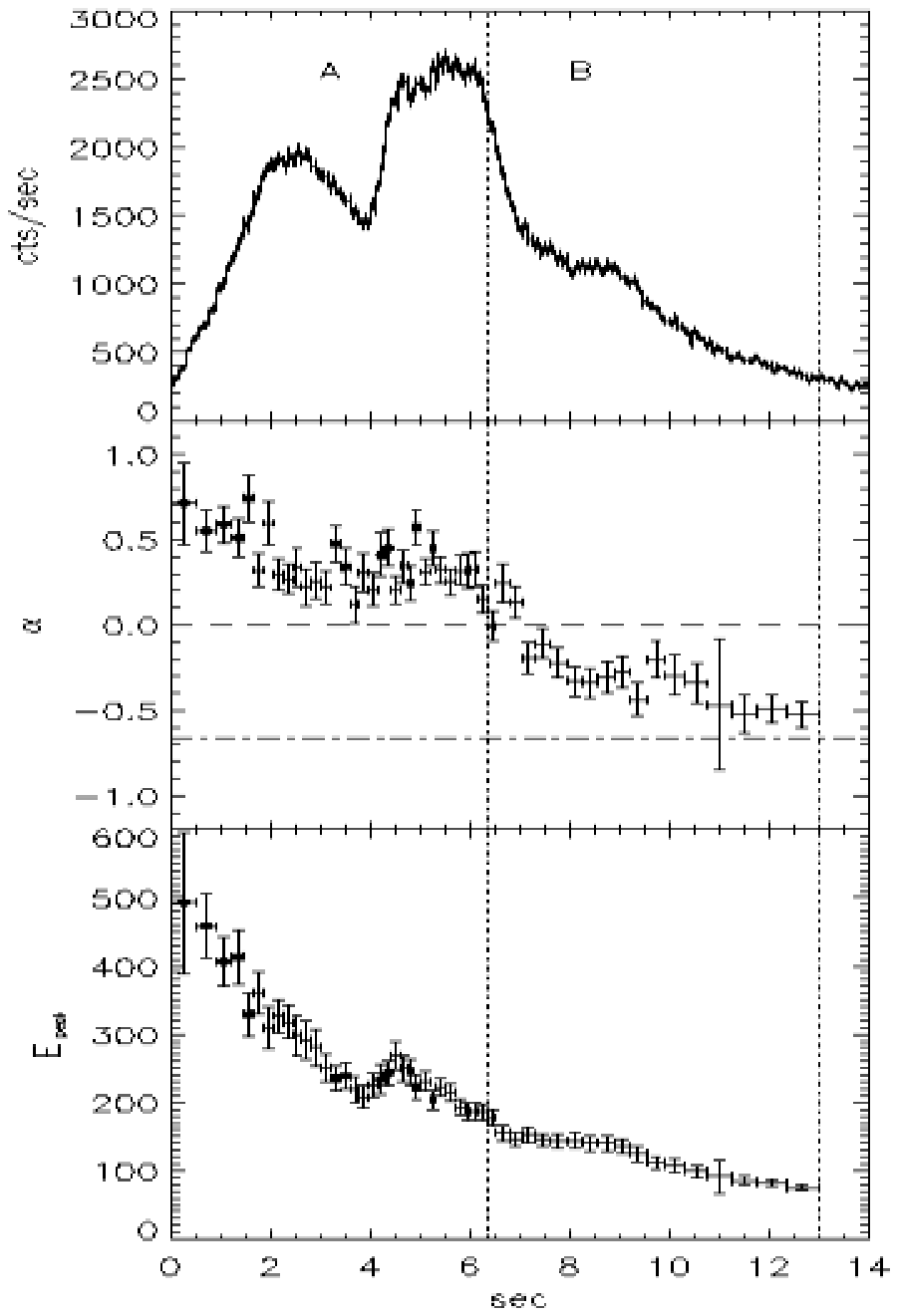}
\caption{ Light curve (top panels) and best fit parameters (photon
 spectral index and peak energy - mid and bottom panel, respectively)
 for GRB 980306 (left) and GRB 911118 (right).  Horizontal lines mark
 the limits $\alpha=-2/3$ and $\alpha=0$.  The filled circles indicate
 the spectra which have been fitted also with a blackbody model.}
\label{ghirlandag1-1}
\end{figure}
\section{Results} 
GRB 980306 represents a newly discovered case of hard burst.  The
spectral evolution of its first 6 s is reported in Fig.1 (left
panel). The evolution of $\alpha$ (mid panel of Fig.1) indicates that,
for most of the pulse, the low energy photon spectrum is harder than
$E^{0}$. The maximum hardness ($\alpha=1.1\pm 0.2$) is reached
at $t \sim 1.5$ s after the trigger.

In GRB 911118 (Fig.1, right panel) the low energy spectral index is
$\alpha \ge 0$ for the first $\sim 6$ s (32 spectra, phase A) The
hardest photon spectrum (at $t\sim 1.5$ s after the trigger) has
$\alpha=0.74\pm 0.13$.

In order to have homogeneous results we also re-analyzed three bursts
(GRB 910807, GRB 910927, GRB 970111) already presented in the
literature for their exceptional low energy spectral hardness (Crider
et al. 1997, 2000; Frontera et al. 2000; Preece 2002). 

Considering the relevance of these 5 GRBs for the comparison with the
models, we also tested the independence of their low energy spectral
hardness from possible sources of systematic error such as the
detector response, the background calculation or the high energy
spectral component. Moreover, we directly verified the low energy
spectral hardness comparing the instrumental spectra with a simulated
flat spectrum. These tests (Ghirlanda et al. 2003) indicate that most
of the time resolved spectra of these bursts are harder than $E^0$ at
more than 3$\sigma$.

As anticipated, we performed fits on the hardest spectra also with a
black body model.  In most of these bursts the initial 1.5 s of their
emission is consistent with a black body with a temperature that
decreases with time as roughly $T_{BB}\propto t^{-1/4}$ starting form $kT \sim
100$ keV (Ghirlanda et al. 2003).

\section{Discussion}
GRB 980306 and 911118, together with GRB 910807, 910927 and 970111
 represent a challange for the proposed emission processes.  Their low
 energy photon spectrum is harder than $N(E)\propto E^{0}$ for a major
 part of the pulse/s reaching a maximum hardness of $E^{1.5}$. These
 results are inconsistent with the optically thin synchrotron model
 which can not account for spectra harder than $E^{-2/3}$ (Katz
 1994). Also its variants like synchrotron self absorption
 (Papathanassiou 1999; Granot, Piran \& Sari 2000), small pitch angles
 of the emitting particles (Lloyd \& Petrosian 2000) or Jitter
 radiation (Medvedev 2001) can produce a spectrum at most as hard as
 $N(E)\propto E^{0}$, which is not compatible with the totality of these
 findings.  Comptonization models (Liang et al. 1997; Ghisellini \&
 Celotti 1999) also have difficulties, although they are consistent
 with very hard spectra in a limited range of energies.

The possible thermal character of the initial phase of all these 5
bursts can be interpreted as the emission of the fireball shells when
they reach the transparency radius (Paczynski 1986; see also Daigne \&
Mockovitch 2002). For typical fireball parameters and using our
observational findings this radius results $R(\tau\sim 1) \sim 5
\times 10^{13}$ cm and the corresponding bulk Lorentz factor is
$\Gamma \sim 1000$. An alternative scenario is the production of an
observed thermal spectrum by the Compton drag of soft (seed) photons
due to the fireball bulk motion (Lazzati et al.  2000). In this case
the seed photons are required to have a temperature of $\sim 5\times
10^{4}$ K in order to produce the observed thermal spectrum. These
results are consistent with the typical values assumed in the thoery
of GRBs (e.g., Rees \& M\'esz\'aros 1994).

At present, we are not able to discriminate between the photospheric
and the Compton drag scenario.  Consider also that emission at times
greater than a few seconds can well be due to other processes,
possibly linked to internal shocks starting to dominate at later
phases.  However, a key difference between the two scenarios is the
fact that the seeds photons for the Compton drag process can be
``used'' only by the first shells, because the time needed to refill
the scattering zone with new seeds exceeds the duration of the burst.
Therefore observing black body emission for a long time, or during the
rising phase of two time resolved peaks of the same GRB would be
difficult to explain in terms of the Compton drag process.

\begin{acknowledgements}
This research has made use of data obtained through the
High Energy Astrophysics Science Archive Research Center
On line Service, provided by the NASA/Goddard Space Flight
Center.  Giancarlo Ghirlanda and AC
acknowledge the Italian MIUR for financial support.
\end{acknowledgements}

\end{document}